\documentclass[a4paper,11pt]{article}
\usepackage{pos}

\title{Quantum Simulations of Fundamental Physics}

\author*[a]{Martin J. Savage}

\affiliation[a]{InQubator for Quantum Simulation (IQuS), Department of Physics, University of Washington, \\
Seattle, WA 98195, USA.}

\emailAdd{mjs5@uw.edu}

\abstract{Simulating the dynamics of non-equilibrium matter under extreme conditions  lies beyond the capabilities of classical computation alone. Remarkable advances in quantum information science and technology are profoundly changing how we understand and explore fundamental quantum many-body systems, and have brought us to the point of simulating essential aspects of these systems using quantum computers. I discuss  highlights, opportunities and the challenges that lie ahead.}

\FullConference{The 11th International Workshop on Chiral Dynamics (CD2024)\\
 26-30 August 2024\\
Ruhr University Bochum, Germany\\}

\begin{document}
\maketitle

\section{Introduction}
\label{sec:intro}
\noindent
Many of the important challenges facing researchers in fundamental science are related to the dynamics of systems far from equilibrium.  
Precision studies  lie beyond the capabilities of HPC alone~\cite{Klco:2021lap,Bauer:2022hpo,Bauer:2023qgm,Beck:2023xhh}, 
limited by “sign problems” in known classical methods, arising from the interference of complex amplitudes. 
Feynman and others~\cite{Feynman:1981tf,5392446,Manin1980,Benioff1980} emphasized the potential of quantum computers for simulating 
fundamentally
quantum systems, in particular, how they would be able to simulate aspects of such systems that are beyond the capabilities of HPC~\cite{Lloyd1073}. 
With corresponding efforts in developing  theoretical frameworks, effective descriptions and models, classical techniques and simulations, quantum algorithms, codes, workflows, and physics-aware optimizations, quantum computers provide a potential path forward for robust simulations of the dynamics of matter initially far from equilibrium or subject to extreme conditions, providing that the relevant initial states can be prepared with sufficient fidelity. 
New understandings of quantum many-body systems (QMBSs) and quantum field theories (QFTs) acquired in these pursuits also accelerate advances in 
quantum information science, technology and engineering
(QISET) and other science domains.

\begin{figure}[h!]
    \centering
    \includegraphics[width=0.75\textwidth]{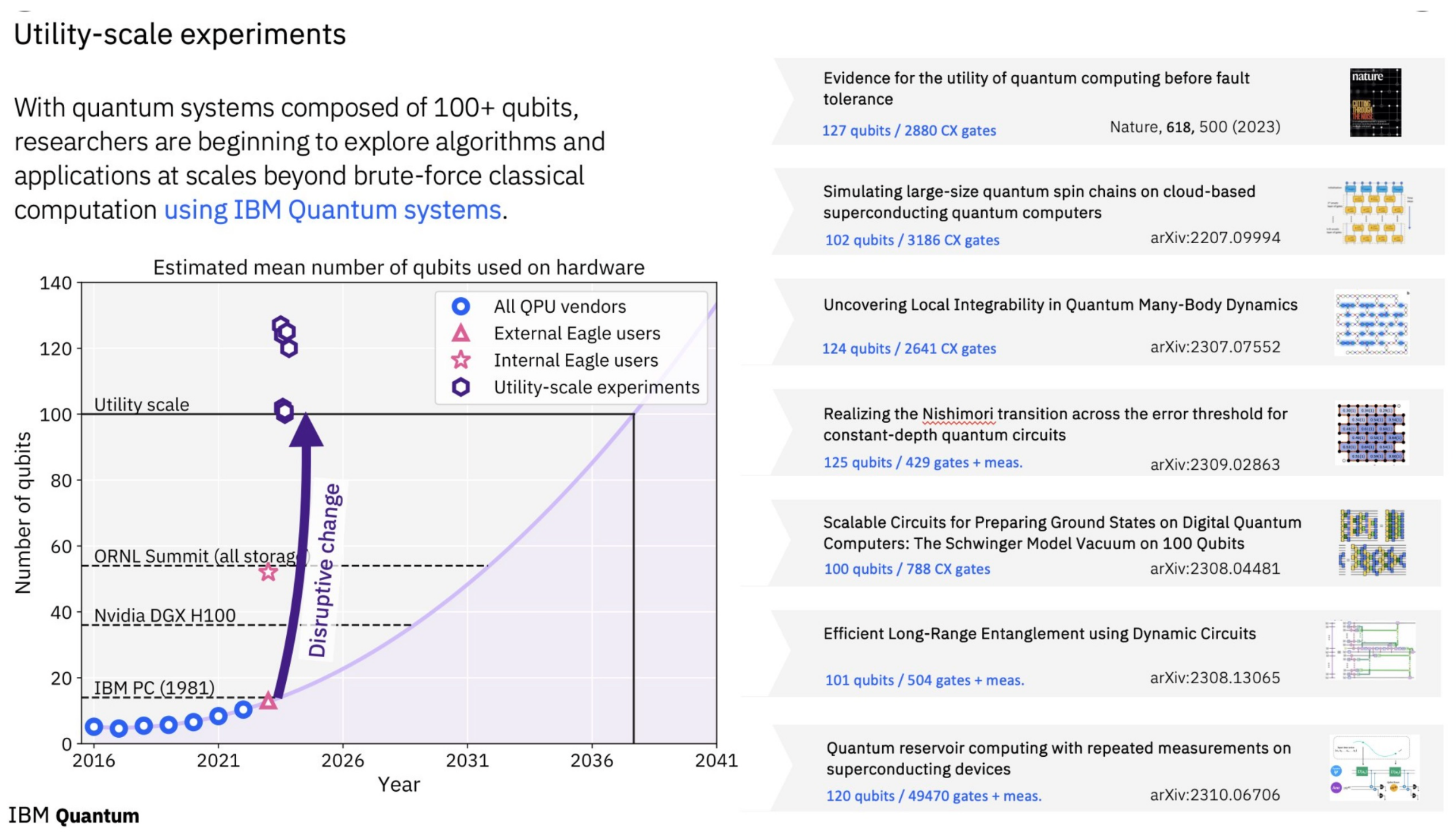}
    \caption{The number of utilized qubits  in jobs run on IBMs quantum computers from 2016 forward~\cite{JayG}.
    Also shown are the classical computers corresponding to the Hilbert-space dimensionality (dashed lines).
    }
    \label{fig:Jay}
\end{figure}

The first digital quantum computers became available to researchers in high-energy physics (HEP) and nuclear physics (NP)
via cloud access in 2017~\cite{IBMQ},  enabling small quantum circuits to be run on (noisy)
5-superconducting qubit quantum processing units (QPUs), 
e.g., Refs.~\cite{Dumitrescu:2018njn,PhysRevA.98.032331}.  
In contrast, 
present day quantum computers range from superconducting-qubit (more than 100 qubits), 
trapped-ion (more than 50 qubits), cold-atom (more than 1000 atoms), and more, with some platforms also supporting qudits.  
Creative algorithms, workflows and implementations are enabling 
simulations of utility using Noisy Intermediate Scale Quantum (NISQ)  computers~\cite{Preskill2018quantumcomputingin}. 
A dramatic illustration of the progress in quantum simulations is shown in Fig.~\ref{fig:Jay}.
One sees that a number of communities have advanced to the point where they are able to perform large-scale quantum simulations, 
i.e., utilizing more than 100 qubits
This happened, in large part, through encouragement 
from technology companies, particularly IBM,
to  ``think big''.  The projects shown in Fig.~\ref{fig:Jay} include 
the preparation of the vacuum in the 
Schwinger model~\cite{Farrell:2023fgd} and time-evolution of wavepackets~\cite{Farrell:2024fit}, accomplished by an IQuS team.
Logical qubits have been demonstrated, 
including from 
IBM using superconducting qubits and Low-Density Parity-Check (LDPC)~\cite{Bravyi_2024},
from the Quantinuum-Microsoft collaboration using trapped-ions~\cite{QuantMicro},
from Atom Computing-Microsoft using cold-atoms~\cite{AtomMicro},
from Google-AI using superconducting qubits and surface codes~\cite{GoogleAILogical}, 
and from
Amazon (AWS) with bosonic cat-codes~\cite{AWScatcodes},
indicating that 
quantum computers with some degree of fault tolerance (FT)
will become available in the near future. 
Mid-circuit measurements that are now possible using some quantum computers 
offer the potential to (significantly) reduce the depth of quantum circuits.

We are in the enviable situation of knowing the underlying interactions and particle content that we wish to simulate, provided by the Standard Model.  While we have made great progress since the 1970's 
in determining masses, energies and low-energy scattering processes using lattice QCD with HPC, 
higher energy dynamics and dense systems of fermions suffer from well-known sign problems.  
Future simulations using quantum computers, with operations that extend beyond the classical gate set,
provide a path forward for addressing 
key aspects of
such systems using the Hamiltonian framework.
Even before fault-tolerant/error-correcting quantum computers become available to provide high-fidelity results, 
much progress can be made along the way from present-day NISQ-era devices.
Asymptotically, we expect to be able to perform end-to-end simulations of complex dynamics, but well before then,
constraints on dynamical aspects of problems, such as, for example,  
fragmentation functions or in-medium energy-loss, will become accessible, and used to refine HPC simulations.

In simulating quantum field theories using quantum computers, pioneering work by Jordan, Lee and Preskill~\cite{Jordan1130,DBLP:journals/qic/JordanLP14,Jordan_2018}
developed a complete protocol for simulating scattering in $\lambda\phi^4$ scalar field theory, 
starting from preparing initial state wavepackets in the interacting theory, 
through time-evolution through the scattering process, and then through to particle detection.
Further, they showed this to be BQP-complete, so that any systems that can be efficiently simulated 
using a quantum computer can be mapped with polynomial-scaling quantum  resources to $\lambda\phi^4$ with external classical sources.

Simulating non-Abelian gauge theories is more complicated than $\lambda\phi^4$ for a number of obvious reasons.
From a practical standpoint, preparing the initial state ``beam'' of hadrons is significantly more challenging, however, that has been essentially achieved in 1+1D using a technique that can be translated to Yang-Mills in 3+1D~\cite{Farrell:2023fgd,Farrell:2024fit}.
The scattering process requires evolving a theory with quark and gluon degrees of freedom forward in time, while maintaining color neutrality despite recent  observations that naive Trotterization induces color-violating amplitudes~\cite{Farrell:2022wyt}.
The digitization of the gauge space is not unique, and current strategies, such as the Kogut-Susskind (KS) Hamiltonian require a substantial Hilbert space per gauge link that grows toward the continuum limit.
An added complication is that applications of the plaquette operator require operations within four gauge spaces of six links
(that have to be re-coupled in 3+1D).   
Recent suggestions of using honeycomb~\cite{Muller:2023nnk,Turro:2024pxu} in 2+1D and triamond~\cite{Kavaki:2024ijd} or 
hyper-honeycomb~\cite{Illa:2025dou} in 3+1D 
somewhat mitigate that complexity.

The last few years has witnessed a growing diversity in quantum computing architectures, 
from trapped-ions with all-to-all connectivity,
cold-atom systems with increasing connectivity, 
to super-conducting systems with nearest-neighbor connectivity.
The atomic systems are characteristically ``slow'' with measurement rates in the Hertz range, 
while superconducting systems are many orders of magnitude faster.
Internationally, there are major efforts toward simulating multiple systems of interest on all available devices. 
At this stage, there are no ``clear winners'' with regard to quantum architectures that will furnish first quantum advantages.

\begin{figure}[h!]
    \centering
    \includegraphics[width=0.8\textwidth]{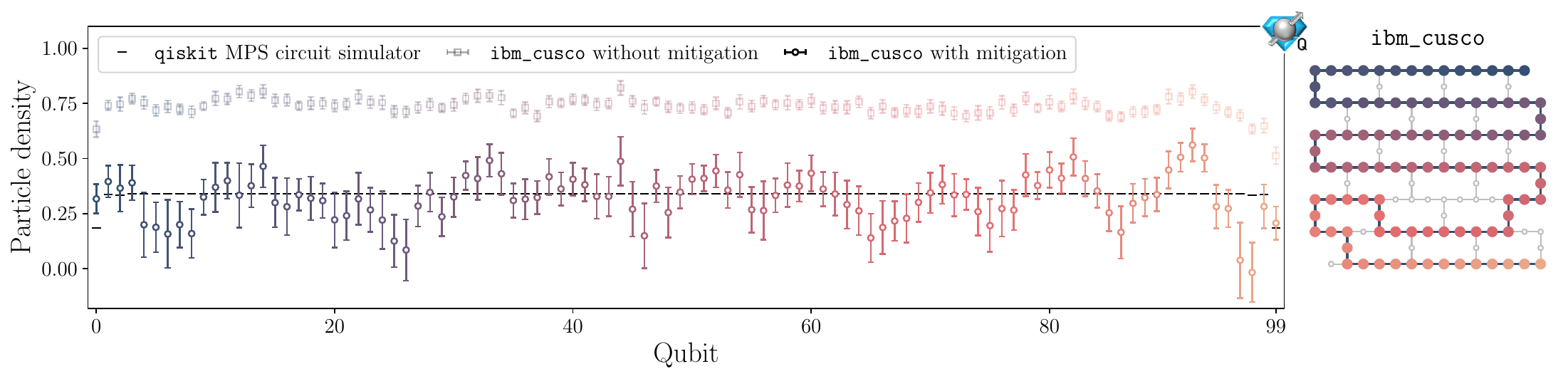}
    \includegraphics[width=0.8\textwidth]{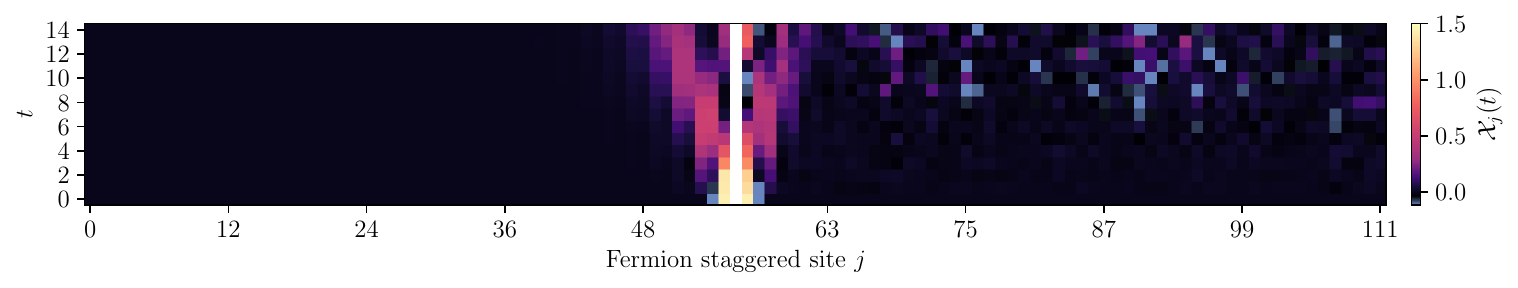}
    \caption{The vacuum of the Schwinger model prepared using IBM's quantum computers, 
    {\tt ibm\_cusco}  (upper panel), and    
    the time evolution of a wavepacket, producing hadrons propagating within the light cone (lower panel).  
    The left side of the lower panel shows the exact result computed using HPC, 
    while the right side shows the results obtained using IBM's quantum computers, {\tt ibm\_torino}~\cite{Farrell:2023fgd,Farrell:2024fit}.
    }
    \label{fig:IBM_SM}
\end{figure}

Low-dimensional models with features in common with 3+1D Standard Model systems of interest have been the focus of 
quantum algorithm and simulation development until now, with only modest efforts to extend to 2+1D and 3+1D.
One of the models that has proven remarkably fruitful is the Schwinger model, which is quantum electrodynamics in 1+1D.
This model confines charges, exhibits a fermion vacuum condensate, and two- and three-body bound states, making it an ideal ``sandbox'' for developing quantum algorithms and intuition for future simulation efforts in quantum chromodynamics (QCD).
Lattice symmetries and confinement led to scalable quantum circuits for preparing the 
quantum vacuum and  wavepackets~\cite{Farrell:2023fgd,Farrell:2024fit}.  
This work built utilized the ADAPT-VQE algorithm~\cite{Grimsley_2019} by working with a  pool
of scalable operators.  
Error mitigation is essential for successful simulations utilizing large quantum volumes~\cite{Kim:2023bwr}.
Figure~\ref{fig:IBM_SM} shows the results of vacuum preparation obtained from IBMs quantum computers 
using scalable operators tuned with small-scale classical simulations and well-known asymptotic behaviors.  
Also shown is the dynamical evolution of a wavepacket into pulses of hadrons moving back-to-back.
There have been a number of important algorithm developmental works using classical computing 
for simulating 1+1D theories using quantum computers, these include kink-kink scattering in ``Ising plus'' spin system to study energy and entanglement evolution~\cite{Milsted:2020jmf,Jha:2024jan},
studies of hadronization and string breaking~\cite{Barata:2023jgd,Florio:2024aix,Farrell:2024mgu}, 
and detailed explorations of algorithm performance for state preparation and evolution in fermion systems, e.g., Refs.~\cite{Chai:2023qpq,Davoudi:2024wyv}.

1+1D simulations of SU(2) and SU(3) theories including 
quarks and anti-quarks, with the gluon field constrained by Gauss's law, have been performed using quantum computers~\cite{Klco:2019evd,PhysRevD.103.094501,Atas:2021ext,Farrell:2022wyt,Farrell:2022vyh,Atas:2022dqm}.
A single spatial lattice site requires $2 n_f n_c$ qubits using the Jordan-Wigner mapping.
These works have included preparing the ground state for a small number of spatial sites using IBMs and Quantuum quantum computers,  examining the time evolution following quenches, and the probing the structure of exotic hadrons.
Some of the new quantum architectures support qudits, e.g., Ref.~\cite{Meth:2023wzd,Nguyen:2023svc,Champion:2024ufp}.  
For example, each of the  trapped ions in some new  systems can support not just two quantum states, but multiple.
We suggested that 
embedding the quarks at each site into
$d=8$ qudits~\cite{Illa:2024kmf}, 
or qu8its, will reduce the depth of quantum circuits due to the capabilities of these devices 
for parallel gate application.
These advances are stimulating the community to consider the advantages and disadvantages of simulating fundamental systems using qubits vs qudits.
First simulations of a highly truncated quantum field theory using a quantum computer 
have been recently performed~\cite{Meth:2023wzd}.

Dynamical quantum phases transitions (DQPTs) can be explored using quantum computers,
opening up new possibilities for understanding fundamental systems.
Interestingly, in the Schwinger model, there is a DQPT arising in the dynamics following a $\theta$-quench, depending on the 
magnitude of $\Delta\theta$~\cite{PhysRevLett.122.050403}.  
Examining the phase of the Fourier components of the two-point function as a function of time reveals a DQPT at $\theta=\pi/2$, above which vortices are generated at critical values of momentum.
This manifests itself as a zero in the Loschmidt echo and non-analytic structure in the associated rate function, 
which is smoothed to some degree in finite systems.  Quantum simulations were performed using IonQ's tapped-ion system, recovering expectations within uncertainties~\cite{Mueller:2022xbg}.

Classical simulations of the QCD phase diagram have been modeled using the Nambu-Jona-Lasinio model with a chemical potential~\cite{Czajka:2021yll}.
The resources of obtaining observables using the Hamiltonian framework  were found to scale reasonably, in contrast to computations in Euclidean space of the same quantities that exhibit sign problems.
A new method for evaluating thermal expectation values was developed, 
working with operations on pure states (called Physical Thermal Pure Quantum (PTPQ)), 
and studied classically in the context of $Z_2$ lattice gauge theory~\cite{Davoudi:2022mbj}.  

While significant progress has been made in simulating 
1+1D  gauge theories, 
the corresponding developments for 2+1D and 3+1D simulations are less advanced~\cite{Klco:2019evd, Ciavarella:2021nmj,Ciavarella:2021lel,ARahman:2021ktn,Illa:2022jqb,ARahman:2022tkr,Ciavarella:2024fzw}.
Formal progress is being made to identify good and hopefully optimal mappings of these theories to qubit and qudit systems with
particular architectures, e.g., Refs.~\cite{Davoudi:2020yln,Bauer:2021gek,Hartung:2022hoz,Grabowska:2022uos,DAndrea:2023qnr,Alexandru:2023qzd,Gustafson:2023kvd,Grabowska:2024emw,Gustafson:2024kym,Halimeh:2024bth}.
For Yang-Mills gauge theories, most of the efforts have been  toward implementing the  KS Hamiltonian, involving chromo-electric and chromo-magnetic field operators, acting on  fabrics of links, each  
 link supporting SU(N) representations.
 At each vertex,  links  are combined to form a gauge-invariant state.
Byrnes and Yamamoto and others  have shown how to map  Yang-Mills to quantum registers~\cite{Byrnes_2006,Zohar:2012xf,Banerjee:2012xg,Tagliacozzo:2012df}.
So far, only 
simulations involving
highly-truncated gauge spaces and spatial volumes have been attempted. 
An exception to this is in recent work by Bauer and Ciavarella who demonstrated significant simplifications by working in the 
large-N$_c$ limit~\cite{Ciavarella:2024fzw}.
Abelian theories have proven less complicated to implement, as expected, and a number of new formulations have appeared, 
typically making explicit use of U(1) gauge invariance and novel choices of gauge fixing~\cite{Haase:2020kaj}.
Small plaquette systems 
have been prepared in the ground state, and also quenched from the trivial vacuum, 
time evolved with the Trotterized KS Hamiltonian.
One ``sociologically-important'' simulation was recently performed to demonstrate the potential for determining transport properties of non-equilibrium matter, in this case the viscosity of finite-temperature SU(2) Yang-Mills gauge theory.
Turro, Ciavarella and Yao~\cite{Turro:2024pxu,Turro:2025sec} 
simulated the real-time correlation function of the stress-energy tensor for a truncated SU(2) gauge theory mapped to a honeycomb lattice~\cite{Muller:2023nnk}.  While this simulation does not provide a quantification of uncertainties, it demonstrates that viscosity is expected to be determined efficiently from quantum simulations, and obtained a result that is consistent with that found in heavy-ion collisions.

Neutrinos play a key role in the evolution of supernova and hence the production of heavy elements in our universe.
Their weak interactions with matter and among themselves combined with an extreme range of energy scales, means that 
classical computing and analytic results alone are limited in their scope of predictive capabilities.
Near the supernova core, 
the extreme density of neutrinos overcomes the weakness of the interactions among them,
rendering $\nu\nu$ interactions the driver for coherent flavor evolution.
Quantum computers are currently enabling studies of the evolution of initial quantum states of neutrinos 
in model systems, both in the spectrum and in dimensional reductions.  
Early simulations using superconducting-qubit and trapped-ion systems examined the quantum properties of systems of less than ten or so neutrinos in simplified models~\cite{Yeter-Aydeniz:2021olz,amitrano23,illa2023multi,illa2022basic}. 
The Hamiltonian describing coherent flavor evolution is derived robustly from the Standard Model, 
which maps to a spin-system (where the spin resides in neutrino flavor space) with interactions among all of the neutrinos.
Interesting properties in the evolution of entanglement was found, including the presence of multi-partite 
entanglement (beyond Bell pairs)~\cite{illa2023multi,Chernyshev:2024kpu}.
These simulations have recently been extended to qutrits, a natural embedding for three flavors of neutrinos~\cite{Turro:2024shh,Spagnoli:2025etu}.
Given the complexity of supernova evolution, there is a long road ahead of development before directly impacting full-scale supernova simulations.

\begin{figure}[h!]
    \centering
    \includegraphics[width=0.55\textwidth]{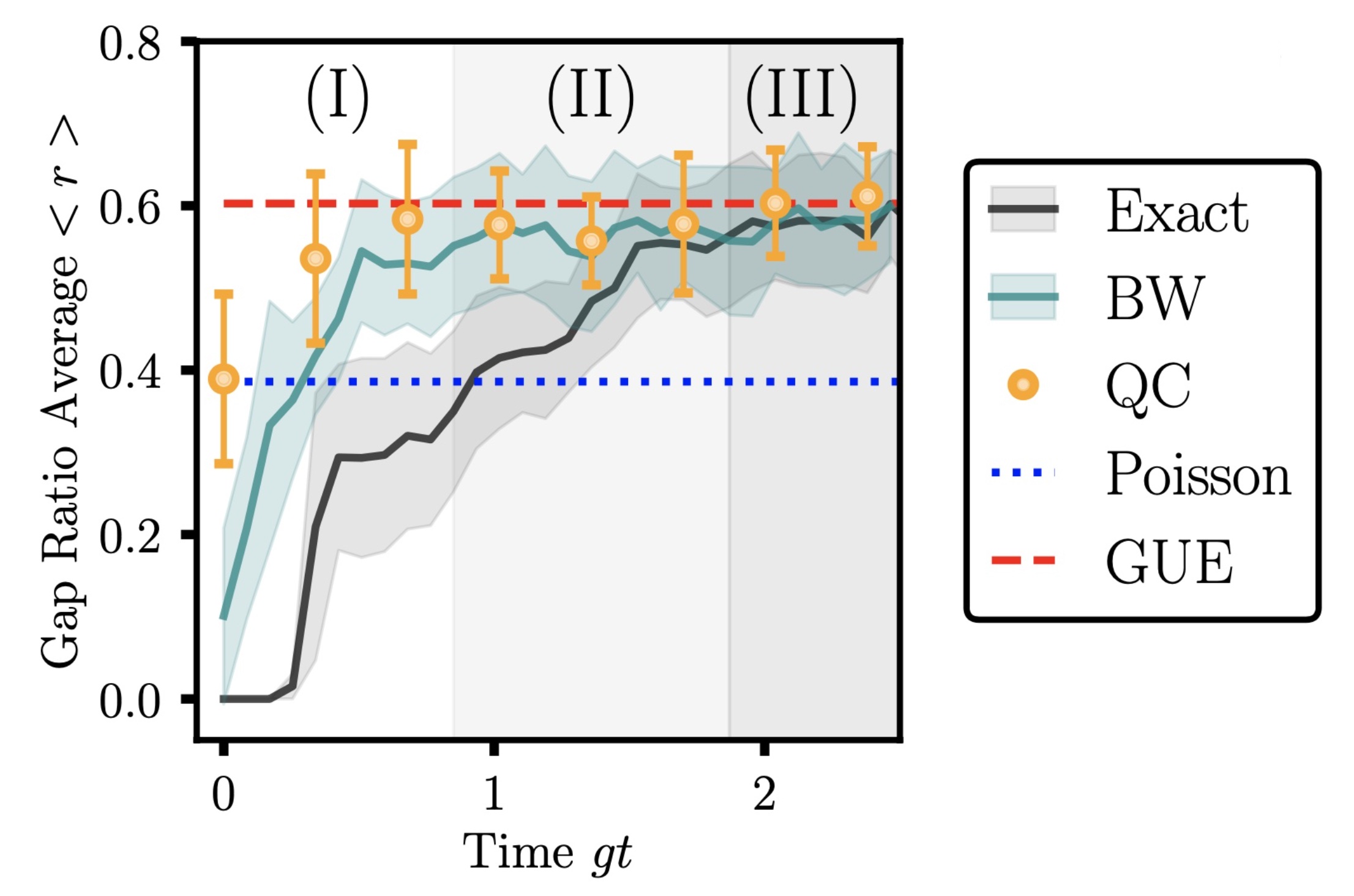}
    \caption{Real-time evolution of the gap-ratio distribution of the entanglement Hamiltonian for a $Z_2$ lattice 
    gauge theory on a dual-rail ring, using 10 plaquettes and 6 randomly selected initial states~\cite{Mueller:2024mmk}.}
    \label{fig:Z2Therm}
\end{figure}

Imagine starting in some initial state, an eigenstate of a given Hamiltonian, 
and then the system is quenched and evolved using a different Hamiltonian.  
Scar states are distributed throughout the evolving system. 
They are only weakly coupled under the evolution and form a ``cold subspace',  delaying thermalization of the system,  and  have low bipartite entanglement entropy.
It was thought until recently that scar states existed only in confining theories, for obvious physical reasons.
However, work by Lewenstein and collaborators~\cite{Aramthottil_2022}, using a $Z_2$ Kitaev model showed that scar states exist in the deconfined phase of this model, and not in the confined phase.  
There have been recent works that examine in detail the entanglement and evolution of entanglement, along with scar states and thermalization in modest-sized, truncated gauge theories.  In particular, $Z_2$ theories mapped to a range of lattice geometries have been examined~\cite{Mueller:2024mmk}.  The theory on the dual-rail ring (or plaquette chain with periodic boundary conditions (PBC)) can be further mapped to a dual theory of single spins per plaquette.   These dual systems are easily amenable to partitioning and detailed studies of entanglement.
Part of this study involved using classical shadows to estimate the reduced density matrix of a sub-partition, and determining eigenvalues of the entanglement Hamiltonian.  Level statistics were determined, and the gap-ratio distribution was seen to evolve from Poissonian at early times to Gaussian Unitary Ensemble (GUE) at late times, indicative of system thermalization,
see Fig.~\ref{fig:Z2Therm}.
Studies of systems closer to physical systems are underway.
There are many studies of thermalization  that have been made possible by progress in quantum computers and algorithms, e.g., Refs.~\cite{Kaufman:2016mif,Zhou:2021kdl}.  
Simulations using cold-atom systems have been performed in which a selection of different pure states
that have the same average energy density  are prepared, and evolved forward in time.   Expectation values of local operators are found to approach the same values at late times, despite following a different evolution path.

One of the very recently emerging areas at the interface of QIS and fundamental physics is ``quantum magic''.  
While this has been a key concept of quantum error correction and communication, it had not penetrated the HEP or NP communities until recently (during an IQuS workshop).  
It is built upon the stabilizer formalism and all of the associated techniques, at the heart of it lies the classical gate set and the universal quantum gate set, and foundational works by Gottesman, Knill and Aaronson~\cite{Gottesman:1998hu,Aaronson_2004}.  
Imagine an n-qubit quantum system initially prepared in the $|0\rangle^{\otimes n}$ tensor-product (classical) pure state.
Acting on this state with an arbitrary selection of gates from the classical gate set, 
comprised of the Hadamard gate, the phase gate and the CNOT gate, produces a stabilizer state.  
This is a state that can be prepared efficiently using classical computation (by definition).  Clearly not all quantum states can be accessed using the classical gate set, which however can be accomplished using a universal quantum gate set - comprised of the classical gates with the addition of the T-gate (as one particular partitioning of gates).  
The magic of a many-body system vanishes for a stabilizer state, by construction, and is a measure of the non-stabilizerness.  The recent introduction of Stabilizer Renyi Entropies (SREs) provide measures that can be used to quantify magic in many-body wavefunctions~\cite{Leone_2022,Haug:2023hcs,Leone:2022lrn,Oliviero_2022}, ${\cal M}_\alpha$,
or in the magic-power of a unitary operator, such as the S-matrix.  
It has been shown through simulation of random quantum circuits, doped with T-gates and single qubit measurements, 
that the transition in sub-system scaling (from volume law to area law) of the entanglement and magic occur in different
regions of doping~\cite{Fux:2023brx}. 
This reinforces the notion that entanglement and magic must be considered as independent measures of the complexity of quantum many-body systems.  This is what follows directly from Gottesman, Knill and Aaronson -- large-scale multi-partite entanglement can be established efficiently with classical computing resources within a given stabilizer state, and thus does not dictate a need for quantum computing resources alone.  Similarly, tensor-product states with large-scale magic can also be prepared efficiently with classical computers using single-qubit rotations.  It is systems with large-scale entanglement and large-scale magic that require quantum computers at scale or for high-precision simulation.  
Recent work quantifying multi-partite entanglement and system-wide magic have been undertaken for p-shell and sd-shell nuclei~\cite{Brokemeier:2024lhq,Robin:2024bdz}, 
and in qubit and qutrit embeddings of neutrinos~\cite{Chernyshev:2024pqy} 
with interesting results.  
For example, the shape-complexity of nuclei, from classical deformation through shape co-existence, through to instability is found to be reflected in measures of magic.

One important area of application in the area of low-energy nuclear physics is nuclear reactions,
many of which are difficult to predict with accuracy.  
While it is a few-body problem, the nature of the nuclear force is such that the structure of nuclei, 
resonating sub-spaces evolving in real-time are beyond the capabilities of classical computing for an array of nuclear reactions,
including those involving short-lived nuclei involved in secondary reactions.  
A co-design team led by Lawrence Livermore National Laboratory is building its own hardware to address these problems~\cite{Turro:2021vbk,Turro:2023dhg}.
Non-relativistic systems have the nice feature that spin and space decouple and reside in distinct Hilbert spaces, and as such, the position of nucleons can be ``handled'' using classical computing, while the spin degrees of freedom, which are intrinsically quantum mechanical, can be ``handled'' using quantum computing, in a hybrid approach~\cite{Turro:2023dhg}.  
Using RF signals design using classical computing to optimally execute requisite quantum operations on their SRF cavity system, this team has successfully simulated the dynamics of two scattering nucleons using this hybrid technique.   Their pulse-control optimizations and hardware designs have enabled simulations of impressive fidelity over relatively long time intervals.

Some of our works have focused on simulating the low-energy 
dynamics in effective models of nucleons, such as the Lipkin-Meshkov-Glick (LMG) model~\cite{Robin:2023pgi,Hengstenberg:2023ryt} 
and the Agassi model (LMG with pairing)~\cite{Illa:2023scc}.
The LMG allows for explorations of  important aspects of many-body systems, including the structure of entanglement~\cite{Hengstenberg:2023ryt}, Hamiltonian learning~\cite{Robin:2023pgi}, 
global optimization for optimal effective model spaces, properties near phase transitions, and more.  With the inclusion of pairing, the Agassi model exhibits super-fluid phases, and 
its intrinsic SO(5) symmetry naturally maps to $d=5$ qudits (qu5its)~\cite{Illa:2023scc}.   
Such a mapping leads to reductions in quantum resource requires for simulations.  
There remains much to be understood about the role of entanglement in nuclei, e.g., Refs.~\cite{gorton18a,Robin:2020aeh,Johnson:2022mzk,Gu:2023aoc,Brokemeier:2024lhq,Robin:2024bdz}

\section{Summary and Outlook}
\noindent
Quantum information science and quantum computers are changing how we think about quantum many-body systems and field theories describing fundamental physics.  
Not only are they changing how we view them, they are changing our objectives for future simulations.  
Significant progress is being made toward quantum computing 
providing predictive capabilities
for the dynamics and properties of matter in extreme conditions beyond what is possible with classical computing and formal techniques alone.  
We have already seen encouraging small-scale simulations of dynamical properties of systems inaccessible 
to classical simulations at scale, but without a complete quantification of uncertainties.

The near future will be a remarkable period in the history of computation as we
will see an evolution from the NISQ-era  to robust fault-tolerant/error-corrected quantum computers. 
Research efforts in NP and HEP, techniques and simulation protocols will need to evolve accordingly, requiring continued engagement among scientists, engineers and developers at universities, technology companies and national laboratories.
We have spent the last eight or so years pursuing and learning from simulations using NISQ-era digital quantum computers, where reliably controlling entanglement through the application of CNOT gates is the major challenge.  This work has led to conceptual paths forward for establishing quantum advantages in some key systems.  
As we evolve into the fault-tolerant/error corrected era,  
optimizing the hybrid classical-quantum processing becomes increasingly important.
This involves physics-awareness and organizing simulations 
in terms of quantum complexity.

\vskip 0.in 
{\it I would like to thank IQuS, my collaborators and the community.  I would also like to thank the organizers of this stimulating and lively meeting.}

\bibliographystyle{JHEP}
\bibliography{refCD2024.bib}


\end{document}